\documentclass{pasj00}
\draft

\begin{document}
\SetRunningHead{Author(s) in page-head}{Running Head}
\Received{2005/05/20}%{2005/03/20}%
\Accepted{2005/09/29}%{2005/09/29}%

\title{HETE-2 Localization and Observations of the Gamma-Ray Burst GRB 020813}

%%% Please use the following style in case that sorting by 
%%% affilation is impossible. 
 \author{
   Rie \textsc{Sato},\altaffilmark{1}
   Takanori \textsc{Sakamoto},\altaffilmark{2}
   Jun \textsc{Kataoka},\altaffilmark{1}
   Atsumasa \textsc{Yoshida},\altaffilmark{4,3}\\
   Motoko \textsc{Suzuki},\altaffilmark{3,1}
   Junichi \textsc{Kotoku},\altaffilmark{1}
   Yuji \textsc{Urata},\altaffilmark{1,3}
   Yoshihisa \textsc{Yamamoto},\altaffilmark{1}\\
   Makoto \textsc{Arimoto},\altaffilmark{1}
   Toru \textsc{Tamagawa},\altaffilmark{3}
   Yuji \textsc{Shirasaki},\altaffilmark{5}
   Ken'ichi \textsc{Torii},\altaffilmark{6}\\
   Masaru \textsc{Matsuoka},\altaffilmark{7}
   Yujin \textsc{Nakagawa},\altaffilmark{4}
   Toru \textsc{Yamazaki},\altaffilmark{4}
   Kaoru \textsc{Tanaka},\altaffilmark{4}\\
   Miki \textsc{Maetou},\altaffilmark{4}
   Makoto \textsc{Yamauchi},\altaffilmark{8}
   Kunio \textsc{Takagishi},\altaffilmark{8}
   Donald Q. \textsc{Lamb},\altaffilmark{9}\\
   Jean-Luc \textsc{Atteia},\altaffilmark{10}
   Roland \textsc{Vanderspek},\altaffilmark{11}
   Carlo \textsc{Graziani},\altaffilmark{9}
   Gregory \textsc{Prigozhin},\altaffilmark{11}\\
   Joel \textsc{Villasenor},\altaffilmark{11}
   J. Garrett \textsc{Jernigan},\altaffilmark{12}
   Geoffrey B. \textsc{Crew},\altaffilmark{11}
   Kevin \textsc{Hurley},\altaffilmark{12}\\
   George R. \textsc{Ricker},\altaffilmark{11}
   Stanford E. \textsc{Woosley},\altaffilmark{13}
   Nat \textsc{Butler},\altaffilmark{11}
   Al  \textsc{Levine},\altaffilmark{11}\\
   John P. \textsc{Doty},\altaffilmark{11}
   Timothy Q. \textsc{Donaghy},\altaffilmark{9}
   Edward E. \textsc{Fenimore},\altaffilmark{14}\\
   Mark \textsc{Galassi},\altaffilmark{14}
   Michel \textsc{Boer},\altaffilmark{10}
   Jean-Pascal \textsc{Dezalay},\altaffilmark{15}
   Jean-Francios \textsc{Olive},\altaffilmark{15}\\
   Joao \textsc{Braga},\altaffilmark{16}
   Ravi \textsc{Manchanda},\altaffilmark{17}
   Graziella \textsc{Pizzichini},\altaffilmark{18}
                        and 
   Nobuyuki \textsc{Kawai}\altaffilmark{1,3}
}
 \altaffiltext{1}{Department of Physics, Tokyo Institute of Technology, Meguro-ku, Tokyo 152-8551}
 \email{rsato@hp.phys.titech.ac.jp}
 \altaffiltext{2}{NASA, Goddard Space Flight Center, Greenbelt, MD 20771 USA}
 \altaffiltext{3}{RIKEN, Hirosawa, Wako, Saitama 351-0198}
 \altaffiltext{4}{Department of Physics, Aoyama Gakuin University, Sagamihara, Kanagawa 229-8558}
 \altaffiltext{5}{National Astronomical Observatory of Japan, Osawa, Mitaka, Tokyo 181-8588}
 \altaffiltext{6}{Department of Earth and Space Science, Graduate School of Science, Osaka University, Toyonaka, \\
                  Osaka 560-0043}
 \altaffiltext{7}{JAXA, Sengen, Tsukuba, Ibaraki 304-8505}
 \altaffiltext{8}{Faculty of Engineering, Miyazaki University, Gakuen Kibanadai Nishi, Miyazaki, Miyazaki 899-2192}
 \altaffiltext{9}{Department of Astronomy and Astrophysics, University of Chicago, Chicago, IL 60637, USA}
 \altaffiltext{10}{Laboratoire d'Astrophysique, Observatoire Midi-Pyr$\acute{e}$n$\acute{e}$es, 14 Ave. Edouard 
                   Belin, \\31400 Toulouse, France}
 \altaffiltext{11}{MIT/CSR 77 Massachusetts Avenue, Cambridge, MA 02139, USA}
 \altaffiltext{12}{Space Sciences Laboratory, University of California, Berkeley, CA 94720-7450, USA}

 \altaffiltext{13}{Department of Astronomy and Astrophysics, University of California at Santa Cruz,\\
                   477 Clark Kerr Hall, Santa Cruz, CA 95064, USA} 
 \altaffiltext{14}{Los Alamos National Laboratory, P.~O. Box 1663 Los Alamos, NM 87545, USA}
 \altaffiltext{15}{Centre d'Etude Spatiale des Rayonnements, Observatoire Midi-Pyr$\acute{e}$n$\acute{e}$es, \\
                   9 Avenue du Colonel Roche, 31028 Toulouse, France}
 \altaffiltext{16}{Instituto Nacional de Pesquisas Espaciais, Avenida Dos Astronautas 1758, \\
                   S$\tilde{a}$o Jos$\acute{e}$ dos Campos 12227-010, Brazil}
 \altaffiltext{17}{Department of Astronomy and Astrophysics, Tata Institute of Fundamental Research, \\
                   Homi Bhabha Road, Mumbai, 400-005, India}
 \altaffiltext{18}{INAF/IASF Sezione di Bologna, via Piero Gobetti 101, 40129 Bologna, Italy}
%% `\KeyWords{}' always has to be placed before `\maketitle'.
 %Do NOT move this preamble from here!

\maketitle

\begin{abstract}

A bright, long gamma-ray burst (GRB) was detected and localized
by the instruments on board the High Energy Transient Explorer 2 
satellite (HETE-2) at 02:44:19.17 UTC (9859.17 s UT) on 2002 August 13.
The location was reported to the GRB Coordinates Network (GCN)
about 4 min after the burst. 
In the prompt emission, the burst had a duration of approximately 125 s, 
and more than four peaks. 
We analyzed the time-resolved 2$-$400 keV energy spectra of 
the prompt emission of GRB 020813 using the Wide Field X-Ray Monitor (WXM) 
and the French Gamma Telescope (FREGATE) in  detail. 
We found that the early part of the burst (17$-$52 s after the burst trigger) 
shows a depletion of low-energy photons below about 50 keV. 
It is difficult to explain the depletion with by either synchrotron self-absorption 
or Comptonization. 
One possibility is that the low-energy depletion may be understood 
as a mixture of ``jitter'' radiation the usual synchrotron radiation component. 

\end{abstract}

\section{Introduction}

%%% about GRB 020813 %%%

It has been widely accepted that gamma-ray burst (GRB) emission is 
produced in a shocked optically thin plasma via the synchrotron process 
(synchrotron shock model; SSM).
The resultant emission spectra is a convolution of synchrotron
emission from electrons with distributed energies, and can be described
as a broken power-law function with a broad $\nu F_{\nu}$ peak.
While the overall spectral shape, particularly at energies around the
peak of the spectral energy density and above, depends critically on the energy
spectrum of the source electrons, the shape of the low-energy spectrum
is bound to converge to a power-law function, $dN/dE\propto E^{-\alpha}$, with
index $\alpha = -0.67$, regardless of the source electron spectrum.  
This is because each of the individual source electron produces emission 
with the same low-energy asymptotic spectral form, 
namely a power-law function with an index of $\alpha = -0.67$.

Most of the observed spectra of GRBs are known to be well
described by the ``Band function'' (Band et al. 1993), which is essentially an
empirical functional form consisting of two smoothly connected power-law
continua, where the spectral index is more negative at higher energies
(Preece et al. 2000).
The Band function has three parameters to characterize the spectral
shape: the low-energy photon spectral index, $\alpha$, the high-energy
photon spectral index, $\beta$, and the break energy, $E_0$.
Here, we use the convention that indices $\alpha$ and $\beta$
are positive  in the power-law function, $dN/dE\propto
E^{\alpha}$, following Preece et al. (1998).
If a photon spectrum produced by the optically thin synchrotron shock
is fitted to the Band function, the low-energy photon spectral index, 
$\alpha$, cannot be larger than $-0.67$, because at any part
of the SSM spectrum the photon slope is more negative than $-0.67$
(e.g., Tavani 1996). 

Preece et al. (1998) examined the time-resolved energy spectra 
of over 100 bright GRBs observed by the BATSE experiment on CGRO, 
and found 23 bursts in which the spectral index limit of the SSM was violated. 
Since the spectral measurement of GRBs at energies below 30 keV was
difficult for BATSE, Preece et al. (1998) used the ``effective spectral
index'' at 25 keV, {\ i.e.} the slope of the power-law tangent to the Band
function at the chosen energy (25 keV).  
Furthermore, Frontera et al. (2000) found several GRBs showing a similar 
low-energy depletion in the range 2--20 keV, 
especially during the initial part of the GRB by the BeppoSAX. 

In this paper we present the spectral observation of GRB 020813 
with the HETE-2 satellite.  
The FREGATE and WXM detectors on HETE-2 cover a wide energy range 
of 2--400 keV, and are well suited to study the spectral shape 
in the X-ray regions.  GRB 020813 has one of the highest spectral peak 
energies, $E_{\rm peak}$, among the GRBs localized by HETE-2 (Sakamoto et al. 2005).  
We found that part of its time-resolved spectra exhibits 
a flat low-energy slope that violates the ``death line'' of the SSM.  
With its long duration, high fluences and characteristic spectra, 
GRB 020813 presents an ideal case to study the low-energy part of a hard GRB.  
Because of its unusual hardness (see figure 4), it is also interesting to study 
the spectral shape at high energies.  
That aspect will be presented in a separate paper using the data 
from another spacecraft, which covers energy ranges higher than those available with HETE-2.  
In section 2, the properties of GRB 020813 observed by HETE-2
are presented, including a detailed analysis of the time-resolved
spectra.  In section 3, we discuss the possible process that causes the
violation of the SSM limit on the low-energy power-law index.  
In particular, synchrotron self-absorption and Compton-scattering processes
are critically examined.
 
\section{Observation}

\subsection{Localization}

The trigger for this event came from the FREGATE instrument, 
at 02:44:19.17 UTC (9859.17 s UT) on 2000 August 13. 
The trigger occurred in the 30$-$400 
keV band, on the 1.3 s time scale.
The WXM flight software localized the burst, and its position 
was RA = $19^{\rm h} 46^{\rm m} 31^{\rm s}$, Dec = $-19^{\circ} 36^{\prime} 27''$ 
(J2000.0)\footnote{Hereinafter, equatorial coordinates are J2000.0.},
with a 90$\%$ error radius of 14$^{\prime}$ which includes 
statistical and systematic errors. 
The flight location was reported in a GRB Coordinate Network (GCN) Position 
Notice at 10:50:48 UT, 4 minutes 14 s after the burst. 
Ground analysis of the WXM data for the burst produced a refined location, 
which was reported in a GCN Position Notice 117 min after the burst. 
The WXM location can be expressed as a 90$\%$ confidence circle 
that is 4$^{\prime}$ in radius and is centered at RA = $19^{\rm h} 46^{\rm m} 41^{\rm s}$, 
Dec = $-19^{\circ} 38^{\prime} 42''$.
Ground analysis of the SXC data for the burst also produced a refined location, 
which was reported in a GCN Position Notice 184 min after the burst. 
The SXC location can be expressed as a 90$\%$ confidence circle 
that is 60$^{''}$ in radius and is centered at RA = $19^{\rm h} 46^{\rm m} 38^{\rm s}$, 
Dec = $-19^{\circ} 35^{\prime} 16''$. 
GRB 020813 was also observed by Interplanetary Network (Ulysses, Konus-Wind, 
and Mars Odyssey) and the resulting IPN localization was fully consistent with 
the WXM error circle (Hurley et al. 2002a, 2002b).

An optical afterglow (OT) was observed 112 min after the burst 
at RA = $19^{\rm h} 46^{\rm m} 41^{\rm s} .88$, Dec = $-19^{\circ} 36^{\prime} 05''.1$ 
(Fox et al. 2002) inside the WXM error circle, 
and follow-up observations were carried out by dozens of telescopes 
distributed around the world. 
The light curve of the optical afterglow has shown a break at 14 hr 
after the burst (e.g., Urata et al. 2003).
In the framework of jetted fireballs, this break corresponds to 
a jet half-opening angle of 1$^{\circ}$.9 $\pm$ 0$^{\circ}$.2 (Covino et al. 2003).
Moreover, optical spectroscopic observations at the Keck observatory 
have determined its redshift as $z = 1.255$ (Barth et al. 2003).
Figure 1 shows the WXM and SXC localizations, which are consistent 
with the location of the OT. 

\begin{figure}
\begin{center}
\includegraphics[width=8cm,angle=270,keepaspectratio]{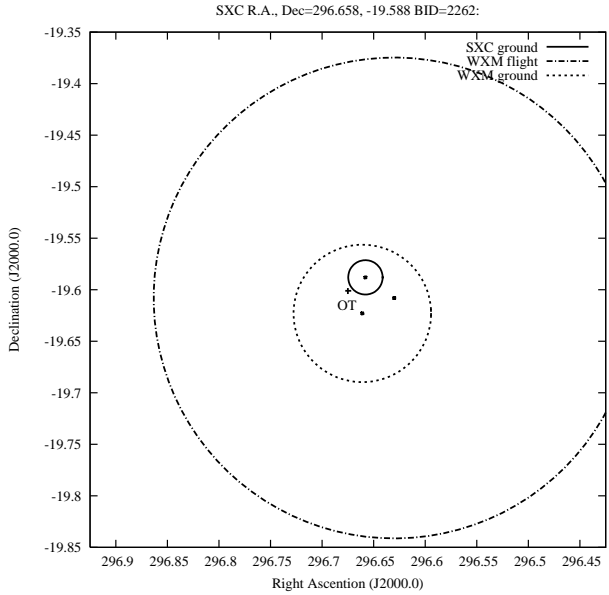}
\end{center}
\caption{Sky map summarizing the localization reported in the GCN burst position notices. The error circles represent 90$\%$ confidence limits.}
\end{figure}

\subsection{Temporal Properties}

Figure 2 shows the complex light curve of GRB 020813 in the four WXM energy bands 
(2$-$5, 5$-$10, 10$-$17, and 17$-$25 keV) and the three FREGATE energy bands 
(6$-$40, 40$-$80, and 32$-$400 keV). 
There are more than three peaks in the WXM light curve 
and at least four spiky peaks in the FREGATE light curve. 
The photon counts in the peaks are higher at later phases.
An inspection of the burst light curve in the WXM and FREGATE energy 
bands shows that the fourth peak is much harder than the others. 

Table 1 gives $t_{50}$ and $t_{90}$ durations of GRB 020813 
in the 2$-$5, 5$-$10, 10$-$25, and 2$-$25 keV  WXM bands and the 
6$-$40, 40$-$80, 32$-$400, and 6$-$400 keV  FREGATE energy bands. 
Moreover, the timescale of the temporal structure seems to be shorter 
at higher energies, which is a feature commonly observed in many GRBs 
(e.g., Fishman et al. 1992, Link et al. 1993, Fenimore et al. 1995). 
We can see a number of ``shots'' in the 32$-$400 keV light curve.

\begin{figure}
\begin{center}
\includegraphics[width=8cm,keepaspectratio]{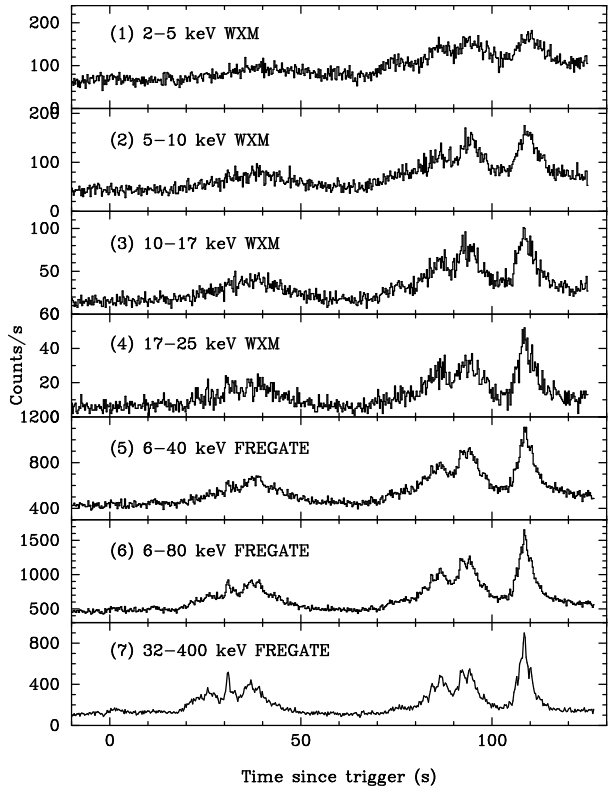}
\end{center}
\caption{Time history of GRB 020813 as observed by the $HETE$-2: 
WXM data (1) 2$-$5 keV, (2) 5$-$10 keV, (3) 10$-$17 keV, and 
(4) 17$-$25 keV bands; and FREGATE data (5) 6$-$40 keV, (6) 40$-$80 keV, 
and (7) 32$-$400 keV bands.}
\end{figure}

\begin{table}
\begin{center}
\caption{Quoted errors corresponding to $\pm 1 \sigma$.}
\begin{tabular}{lccc}
\hline\hline
Instrument & Energy  & $t_{90}$  & $t_{50}$ \\
           & (keV)   & (s)       & (s)      \\
\hline
HETE-2 WXM & 2$-$25  & $121.6\pm1.2$ & $34.4\pm0.1$ \\ 
             & 2$-$5   & $129.0\pm1.2$ & $38.1\pm1.3$ \\
             & 5$-$10  & $119.1\pm2.2$ & $33.2\pm1.2$ \\
             & 10$-$25 & $99.5\pm1.8$ & $34.4\pm1.7$ \\
HETE-2 FREGATE & 6$-$400 & $87.9\pm1.1$ & $65.5\pm0.8$ \\
             & 6$-$40   & $85.9\pm0.3$ & $32.7\pm0.4$ \\
             & 40$-$80  & $89.0\pm0.3$ & $66.1\pm0.7$ \\
             & 32$-$400 & $88.7\pm0.6$ & $67.3\pm1.0$ \\
\hline
\multicolumn{4}{@{}l@{}}{\hbox to 0pt{\parbox{180mm}{\footnotesize
     }\hss}}
\end{tabular}
\end{center}
\end{table}

\subsection{Spectrum}

%%% FREGATE data %%%
Two types of data sets  (``$burst$ data'' and ``$continuous$ data'') are provided 
by the FREGATE detector on board HETE-2 (Atteia et al. 2003). 
The burst data are only available when FREGATE triggered on the GRB, whereas 
continuous, monitoring data are always recorded whenever HETE-2 is 
operating. In a spectral analysis of FREGATE data, we usually use 
burst data, since it provides time- and energy-tagged photons with a much finer time resolution 
(6.4 $\mu$s) than the continuous data, which is accumulated every 5 s. 
However, due to the long duration of GRB 020813, the memory was full 
and burst data were only available at 0$\le t\le$70 s. 
Therefore, we constructed the energy spectrum for the remaining part of the GRB 
by using continuous data over the full energy range of FREGATE (7$-$400 keV), 
which was also discrete data and only available during three separate intervals: 
$-120\le t\le -40$ s, 40$\le t\le$120 s, and 220$\le t\le$300 s (figure 3, left).
Note that the differences between the $burst$ data and the $continuous$ data are 
its time resolution and energy resolution. 
The quality of the derived spectra is exactly the same for both types of data 
in the following analysis.

The background level at an arbitrary time was estimated by interpolating 
the $pre$-burst $continuous$ data ($-120\le t\le -40$ s; bg1) 
and the $post$-burst $continuous$ data (220$\le t\le$300 s; bg2) with a linear function 
of ${\rm BG \hspace*{0.2cm} count} = 1.13\times10^{4} - 0.87 \times t$, 
where $t$ is the time after the burst in seconds (figure 3, $right$). 
We then calculated a weighted mean of the bg1 and the bg2 spectra at each time $t$, 
and subtracted this from the FREGATE spectral data.

\begin{figure}
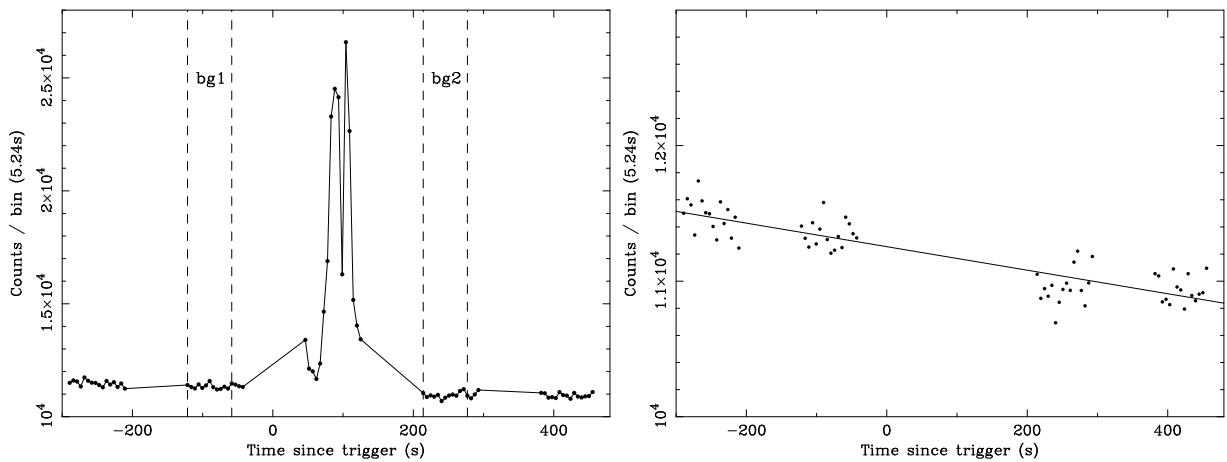

\begin{center}
\includegraphics[width=6cm,angle=270,keepaspectratio]{fig3.ps}
\includegraphics[width=6cm,angle=270,keepaspectratio]{fig4.ps}
\end{center}
\caption{$Left:$ Light curve of the $continuous$ data in the FREGATE 
energy band.
$Right:$ Background of the $continuous$ data with a linear function.}
\end{figure}

%%% WXM data %%%
To analyze the WXM data, we used the ``TAG data'' generated at the burst 
trigger time in the full 2$-$25 keV energy range.  
The background was integrated over 110 s, which was extracted 
from 118 s to 8 s before a burst. In the WXM spectral analysis, 
we considered only those events that registered on wires in the
X- and the Y-detectors illuminated by the burst. Furthermore, 
since the variation in the gain is not uniform at the ends of the 
wires in the WXM detectors (Shirasaki et al. 2003),  we used only the 
photon counts that registered in the central $\pm50$ mm region of the 
wires to construct reliable spectra of the burst. 

As one can see in figure 2, the light curve of GRB 020813 shows a very
complex structure, such that four distinct peaks, at least, are visible 
in the light curve. 
Considering the fast time variability, we analyzed the spectrum of GRB 020813 
every 5 s, which is also the limit imposed by the time resolution 
of the FREGATE continuous data. 

%%% hardness ratio %%%
Before going into the detailed spectral analysis, we first review the 
spectral evolution during a burst, by comparing the photon counting 
rates in the low energy (WXM) band with those in  the high energy (FREGATE) 
bands. The hardness ratio of the fluxes, $S_{\gamma}$(30$-$400 keV) to 
$S_{\rm X}$(2$-$30 keV), is shown in figure 4 (third panel). 
One can see that GRB 020813 possibly consists of two distinct bursts 
that may have different physical origins. The border between the first 
and second bursts is approximately given by $t\sim 60$ s. Note that 
this corresponds to the epoch when the first gradual burst decayed to the 
initial (background) level, and succeeding flares started to appear 
in the light curve (figure 2).

We therefore denote the former part as ``P1'' and the latter part 
of the flare as ``P2'', just for convenience. The time intervals  
for P1 and P2 correspond to 17$-$52 s and 67$-$109 s 
from the burst trigger time, respectively. Note that the peak of P1 
shows larger values of $S_{\gamma}$(30$-$400 keV)/ $S_{\rm X}$(2$-$30 keV) 
than that of P2, meaning that energy spectra in the peak of P1 are harder 
than those of P2.

%%% fitting model %%%
In a more detailed analysis, we considered three different models to
reproduce the observed spectra: (1) a power-law function, 
(2) a cutoff power-law function and (3) a Band function (Band et al. 1993). 
We first fitted the data with a single power-law function, but this
model did not represent the spectra well, except in low photon 
statistics regions. 
This is because the spectrum of GRB 020813 is not a simple power-law, 
but  bends sharply at high energies, as is often reported in other GRBs 
(e.g., Band et al. 1993).  On the other hand, both the cutoff power-law and 
the Band function provide improved fits for all regions. Due to the limited 
photon statistics, it is hard to say which model is a better representation 
of the GRB 020813 spectra for each 5 s segment. We thus fit the time-averaged 
spectrum with a cutoff power-law and the Band function, and compared 
their results. When the photon statistics are sufficiently high, the 
Band function gives a better fit with a reduced $\chi^{2}$ of 1.18 for 
140 dof, compared to 1.33 for 141 dof for the cutoff power-law model. 
In the following analysis, we therefore use the Band function to 
discuss the evolution of the spectra for all time intervals.

%%% result %%%
Table 2 presents the results of a time-resolved spectral analysis 
of GRB 020813. We used the XSPEC version 11.2.0 software package 
to do the spectral fits. The photoelectric absorption in the direction 
of GRB 020813 is 7.0 $\times$ $10^{20}$ cm$^{-2}$ (Vanderspek et al. 2002), 
which is negligible even in the WXM energy range above 2 keV. 
Therefore, we do not consider $N_{\rm H}$ as a parameter in the spectral analysis. 
Figure 4 summarizes the time evolution of spectral parameters together 
with the WXM and the FREGATE light curves measured at 2$-$25 keV 
and 7$-$400 keV, respectively (the first and the second panels). 
The time variations of the low and high-energy photon indices, 
$\alpha$ and $\beta$, are plotted in the fourth and fifth panels. 
The break energy in the spectrum $E_{0}$ is given in the sixth panel, and the fluxes 
measured in the 2$-$400, 2$-$25 and 30$-$400 keV energy bands are plotted 
in the remaining panels.

As shown in figure 4, GRB 020813 spectra have a clear break around 100 keV, 
meaning that most of the power is emitted in the hard X-ray 
and the soft gamma-ray energy bands. For the time-averaged spectrum of P1, 
we found that  
$S_{\rm X}(2-30 \hspace*{0.2cm} {\rm keV})= (2.80 \pm 0.06) \times 10^{-6}$ ergs cm$^{-2}$ 
and 
$S_{\gamma}(30-400 \hspace*{0.2cm} {\rm keV}) = (2.53 \pm 0.07) \times 10^{-5}$ ergs cm$^{-2}$. 
Thus, the logarithmic  fluence ratio, defined as log$(S_{\rm X}/S_{\gamma})$, 
is $-$0.96. 
Similarly, we obtain log$(S_{\rm X}/S_{\gamma}) = -0.69$ for P2, 
with $S_{\rm X}(2-30 \hspace*{0.2cm} {\rm keV}) = (8.20 \pm 0.14) \times 10^{-6}$ ergs 
cm$^{-2}$ and $S_{\gamma}(30-400 \hspace*{0.2cm} {\rm keV})= (3.99 \pm 0.07) 
\times 10^{-5}$ ergs cm$^{-2}$. These results confirm that 
GRB 020813 belongs to the class of long classical GRBs 
according to the HETE classification method 
(Sakamoto et al. 2005). 

However, we  note that log$(S_{\rm X}/S_{\gamma}) = -0.96$ is quite large 
compared to other GRB populations (Sakamoto et al. 2005) and would be at the extreme 
end of the classical GRB population. This is also indicated by the 
unusually $depletion$ of low-energy (below $E_{0}$) photons observed in P1. 
In fact, the low-energy power-law indices, $\alpha$, range from 
$-1.16$ to 0.0, some of which are much larger than expected from the ``death line'' 
($\alpha = -0.67$).
We will thus consider the origin of this low-energy depletion in the next section.

\begin{table}
\begin{center}
\caption{Spectral model parameters of the Band function fit to the time-resolved spectra of GRB 020813.}
\begin{tabular}{rrrrr}
\hline
\hline
\footnotesize{Time interval (s)} & \footnotesize{Photon index} & \footnotesize{Photon index} 
& \footnotesize{Cutoff energy} & \footnotesize{Reduced $\chi^{2}$} \\
\footnotesize{mid-time (start--end)}
& \footnotesize{$\alpha$ ($E^{\alpha}$)} & \footnotesize{$\beta$ ($E^{\beta}$)}
& \footnotesize{($E_{0}$ keV)}
& \footnotesize{(dof)}\\
\hline
7.0  (-0.7 -- 14.8) & 0.6$^{+1.3}_{-0.8}$  & -2.3$^{+0.5}_{-7.7}$ &  44$^{+49}_{-20}$   & 1.0 (18)\\
17.4 (14.9 -- 19.9) & -0.0$^{+5.0}_{-1.0}$ & -1.5$^{+0.4}_{-8.6}$ &  40$^{+828}_{-37}$  & 1.2 (27)\\
22.4 (19.9 -- 24.9) & -0.2$^{+0.2}_{-0.2}$ & -1.8$^{+0.3}_{-8.2}$ & 118$^{+62}_{-30}$   & 1.1 (35)\\
27.7 (25.1 -- 30.2) & -0.3$^{+0.1}_{-0.1}$ & -2.0$^{+0.3}_{-8.0}$ & 106$^{+13}_{-21}$   & 0.8 (47)\\
32.8 (30.2 -- 35.4) & -0.5$^{+0.1}_{-0.1}$ & -1.9$^{+0.2}_{-0.5}$ & 115$^{+23}_{-20}$   & 1.3 (57)\\
38.1 (35.5 -- 40.7) & -0.5$^{+0.1}_{-0.1}$ & -9.4$^{+6.8}_{-0.6}$ &  95$^{+5}_{-9}$     & 1.2 (58)\\
43.6 (41.0 -- 46.2) & -0.7$^{+0.1}_{-0.1}$ & -9.4$^{+7.2}_{-0.6}$ &  65$^{+6}_{-5}$     & 0.9 (45)\\
48.7 (46.2 -- 51.3) & -0.9$^{+0.2}_{-0.2}$ & -3.1$^{+0.8}_{-7.0}$ &  45$^{+18}_{-13}$   & 0.8 (29)\\
53.9 (51.3 -- 56.5) & -1.2$^{+0.3}_{-0.3}$ & -9.4$^{+19.4}_{-0.6}$ & 59$^{+85}_{-31}$   & 0.8 (30)\\
59.4 (56.8 -- 62.0) & -1.1$^{+0.6}_{-0.3}$ & -9.4$^{+19.4}_{-0.6}$ & 42$^{+29}_{-28}$   & 1.5 (30)\\
64.7 (62.0 -- 67.3)\footnotemark[$*$] &   ---    &    ---    &        ---          &  ---    \\
69.9 (67.3 -- 72.5) & -1.4$^{+0.1}_{-0.1}$ & -9.3$^{+7.5}_{-0.7}$ &  87$^{+188}_{-38}$  & 0.8 (30)\\
75.2 (72.5 -- 77.7) & -1.2$^{+0.1}_{-0.1}$ & -9.4$^{+7.4}_{-0.6}$ & 138$^{+62}_{-22}$   & 0.8 (74)\\
80.4 (77.7 -- 83.0) & -0.9$^{+0.1}_{-0.1}$ & -1.8$^{+0.1}_{-0.1}$ &  57$^{+22}_{-17}$   & 1.0 (74)\\
85.7 (83.0 -- 88.2) & -0.9$^{+0.1}_{-0.1}$ & -1.7$^{+0.1}_{-0.2}$ & 138$^{+33}_{-29}$   & 1.2 (73)\\
90.9 (88.2 -- 93.5) & -0.8$^{+0.1}_{-0.1}$ & -1.9$^{+0.1}_{-0.2}$ & 109$^{+15}_{-14}$   & 1.2 (73)\\
96.1 (93.5 -- 98.7) & -0.8$^{+0.1}_{-0.1}$ & -1.8$^{+0.1}_{-0.1}$ &  80$^{+18}_{-16}$   & 0.8 (73)\\
101.4 (98.7 -- 104.0) & -1.2$^{+0.1}_{-0.1}$ & -9.4$^{+7.2}_{-0.6}$ & 111$^{+27}_{-11}$ & 0.9 (68)\\
106.6 (104.0 -- 109.2)& -0.8$^{+0.1}_{-0.1}$ & -1.6$^{+0.1}_{-0.1}$ & 100$^{+26}_{-24}$ & 1.3 (73)\\
\hline
\multicolumn{5}{@{}l@{}}{\hbox to 0pt{\parbox{180mm}{\footnotesize
       \par\noindent
       \footnotemark[$*$] The spectral parameters cannot be determined for the time interval 62.0--67.3 due to poor statistics.
     }\hss}}
\end{tabular}
\end{center}
\end{table}

\begin{figure}
\begin{center}
\includegraphics[width=8cm,keepaspectratio]{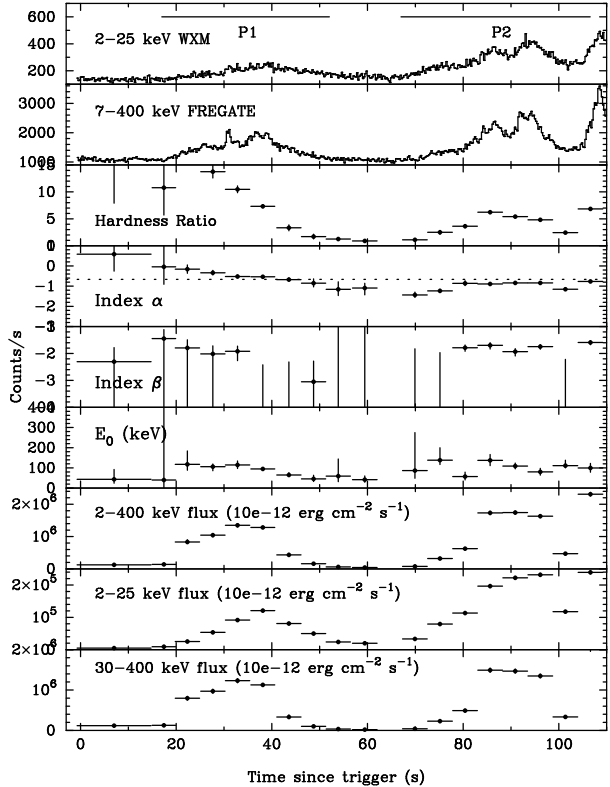}
\end{center}
\caption{Light curve and time evolution of the Band function parameters.}
\end{figure}
 
\section{Discussion}

\subsection{Origin of Low-Energy Depletion in P1}

As we have discussed in the previous section, GRB 020813 shows 
a depletion of low-energy photons, especially in the P1 region. 
Such flat spectra, exceeding the ``death line'' 
($\alpha = -0.67$; Preece et al. 2000), have already been reported 
in a number of BATSE (Burst and Transient Source Experiment) GRBs, 
though they were not conclusive due to the limited sensitivity 
of BATSE at low energies. 
Frontera et al. 2000, using the BeppoSAX WFC and GRBM, reported that a few GRBs 
show significant low-energy depletion in 2--20 keV, though their energy coverages 
of WFC and GRBM are not completely continuous, and the energy resolution of GRBM 
was somewhat limited. 
Therefore, the HETE-2 observation of GRB 020813 presents an independent measurement 
of a similar characteristic spectra with higher reliability over the wide and continuous 
energy range (2--400 keV) of HETE-2. 
In the following, we examine various scenarios to understand the origin 
of the X-ray emission from GRB 020813. 

First, we consider the spectral hardening due to synchrotron self-absorption (SSA). 
We can define a critical frequency, $\nu_{\rm a}$, where
the plasma becomes optical thick, resulting in a hard X-ray 
spectrum with $\alpha$ = $+1$ well below $\nu_{\rm a}$ (e.g., Sari et al. 1998).  
Meanwhile, the high-energy spectral index 
of GRB 020813, $\beta \sim -2.15$, is that expected from a ``fast
cooling'' of electrons with its power-law index, $-2.3$ (= 2$\beta$+2).
Therefore, a sharp spectral break, as observed in GRB 020813, is expected 
only when the minimum synchrotron frequency, $\nu_{\rm m}$, 
is $degenerate$ with respect to $\nu_{\rm a}$, such that  $h\nu_{\rm m} 
\sim h\nu_{\rm a} \sim E_{0}$.

In this particular case, the observed SSA frequency is approximately given as
\footnote{Using an approximate relation of $\delta \sim \Gamma_{\rm BLK} 
\sim \gamma_{\rm m}$, where $\delta$ is the Doppler beaming factor, 
$\gamma_{\rm m}$ is the minimum electron Lorentz factor 
and $\Gamma_{\rm BLK}$ is the bulk Lorentz factor. 
Assuming a viewing angle between the GRB and observer 
$\theta \sim \frac{1}{\Gamma_{\rm BLK}}$, we can derive $\delta \sim \Gamma_{\rm BLK}$. 
Furthermore, we assume $\gamma_{\rm m}$/$\Gamma_{\rm BLK}$ $[$see, equation (1)$]$ 
is of order unity as often assumed in modeling the GRB emission.}

\begin{equation}
\nu_{\rm a} \sim 14.5 \times (1+z)^{-1} (rn)^{3/5} B^{2/5}
        \sim 6.4 \times (rn)^{3/5} B^{2/5} 
\hspace*{0.2cm} {\rm Hz},
\end{equation}
where $z$ is the redshift, $r$ (${\rm cm}$) the radius of emission volume, 
$n$ (${\rm cm^{-3}}$) the electron number density, and 
$B$ (gauss) the magnetic field strength (Dermer et al. 2000). 
Then, the peak frequency of the synchrotron radiation in the 
observer's frame is given as 

\begin{equation}
\nu_{\rm m} \sim 1.2 \times 10^{6} (1+z)^{-1} B \delta \gamma_{\rm m}^{2}
        \sim 5.3 \times 10^{5} B \delta \gamma_{\rm m}^{2}
\hspace*{0.2cm} {\rm Hz}
\end{equation} (Rybicki, Lightman 1979). 
Combining equations (1) and (2) to eliminate B, and normalizing to typical values 
for GRBs, the electron number density is given by 

\begin{equation}
n =  3.0 \times 10^{15} 
\biggl(\frac{r}{10^{10}} \biggr)^{-1} 
\biggl(\frac{\nu_{\rm a}}{10^{19}} \biggr)^{5/3}
\biggl(\frac{\nu_{\rm m}}{10^{19}} \biggr)^{-2/3} 
\biggl(\frac{\gamma_{\rm m}}{100} \biggr)^{2} 
\hspace*{0.2cm} {\rm cm^{-3}}.
\end{equation}

In figure 5, we plot the allowed parameter space for the region size $r$ 
and the electron number density $n$ of GRB 020813. 
Here, we have assumed $\nu_{\rm a}$ = $\nu_{\rm m}$ = $10^{19}$ Hz. 
Equation (3) is represented by line (I).

Then, from the 
electron energy density, $u_{\rm e} \simeq n \gamma_{\rm m} m_{\rm e} c^2$, 
and the magnetic field energy density, $u_{\rm B} = B^2/8 \pi$, 
with $u_{\rm e} \equiv \xi u_{\rm B}$,  
we obtain

\begin{equation}
n = 1.7 \times 10^{17} \xi
\biggl(\frac{\nu_{\rm m}}{10^{19}} \biggr)^{2}
\biggl(\frac{\gamma_{\rm m}}{100} \biggr)^{-7}
\hspace*{0.2cm} {\rm cm^{-3}}. 
\end{equation}
Equations for the cases of $\xi =$ 0.1, 1, and 10 are represented 
by lines (II) in figure 5.

Finally, from the relation 
 $\frac{4}{3} \pi r^{3} u_{\rm e} = \epsilon_{\rm e}E_{\rm tot}$, 
we obtain 

\begin{equation}
n = 2.9 \times 10^{24} 
\biggl(\frac{r}{10^{10}} \biggr)^{-3} 
\biggl(\frac{\gamma_{\rm m}}{100} \biggr)^{-1}
\biggl(\frac{\epsilon_{e}}{0.1} \biggr)
\biggl(\frac{E_{\rm tot}}{10^{52}} \biggr)
\hspace*{0.2cm} {\rm cm^{-3}},
\end{equation}
where $\epsilon_{\rm e}$ is the fraction of the shock energy given 
to the electrons and $E_{\rm tot}$ is the total explosion energy of GRB 020813,
 $1.2 \times 10^{53}$ ergs (2$-$400 keV), calculated from the 
observed fluence and duration of P1.
Equations for the cases of $\epsilon_{\rm e} =$ 0.001, 0.1, and 0.5 are 
represented by lines (III) in figure 5.

As can be seen in figure 5, the most likely parameters for the SSA model 
would be a region defined by the crossing of three different lines. These are 
$n \sim 4.5 \times 10^{12}$ cm$^{-3}$, $r \sim 1.4 \times 10^{14}$ cm,
$\gamma_{\rm m} \sim 450$, and $B \sim 2.1 \times 10^{5}$ gauss, 
assuming $\xi = 1$ and $\epsilon_{\rm e} \sim 0.1$.

Using these parameters, the peak flux is 
$\sim 10^{9}$ ergs cm$^{-2}$ s$^{-1}$, calculated from 
$\nu F(\nu) = \frac{\nu}{4 \pi d_{\rm L}^2} 
\int d^3 \vec{r} 
[4\pi j_{\nu} (\vec{r}) exp (-\int \alpha_{\nu}(\vec{r^{\prime}}ds^{\prime})]$, 
where $j_{\nu}$ is the synchrotron emission coefficient and
$\alpha_{\nu}$ is the self-absorption coefficient (e.g., Rybicki, Lightman 1979).
It is more than 15 orders of magnitude different from the observed 
value, $10^{-6}$ ergs cm$^{-2}$ s$^{-1}$.
In other words, in order to observe $\nu_{\rm a}$ in the hard X-ray 
band, we have to choose unusual parameters, such as 
$\epsilon_{e}$ $\ll$ $10^{-10}$,  in contrast to the standard 
GRB fireball model (e.g., Piran 1999).

\begin{figure}
\begin{center}
\includegraphics[width=8cm,angle=270,keepaspectratio]{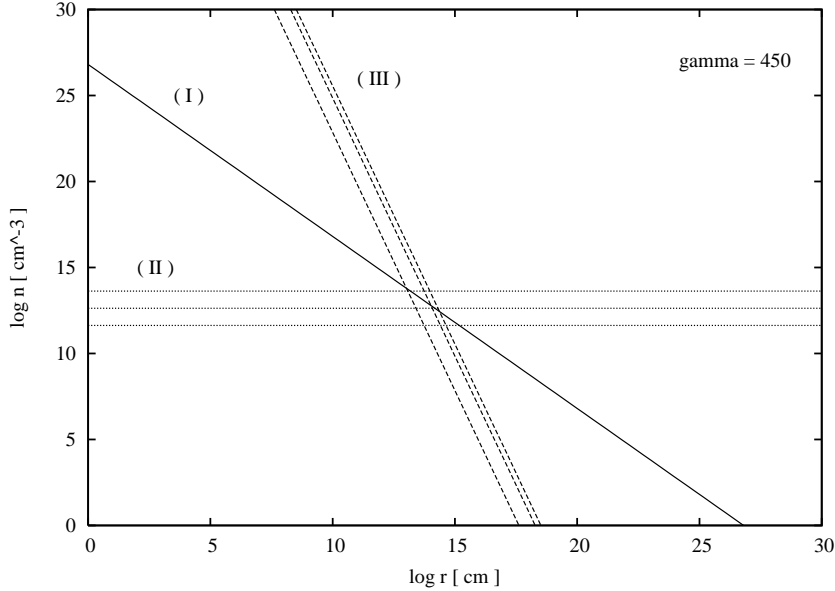}
\end{center}
\caption{Parameter space ($n, r$) allowed by the SSA model with
$\nu_{\rm a} = \nu_{\rm m}$ of $10^{19}$ Hz
and $E_{\rm tot}$ of $1.2 \times 10^{53}$ ergs.
Line (I) was derived from the SSA frequency.
Lines (II) were derived from relation of the energy density of the electrons and the magnetic field.
Here, we used $\xi =$ 0.1, 1, and 10 (from bottom to top).
Lines (III) were derived from a calculation of the total energy of the electrons.
Here, we used $\epsilon_{e} =$ 0.001, 0.1, and 0.5 (from left to right).
The crossing point of the three expressions indicated the allowed parameters for the SSA model.}
\end{figure}

As another possible scenario, we consider the special case that 
the hard X-ray spectrum is the low-energy end of the inverse 
Compton emission, which could be a ``mirror'' of the synchrotron 
self-absorption spectrum in the radio band (e.g. Liang et al. 1997). 
For this case, the synchrotron emission spectrum is boosted by a factor 
of $\gamma_{\rm m}^{2}$ via inverse Compton scattering of synchrotron 
photons. Therefore, we assume $\gamma^{2} \nu_{\rm a} \equiv
\nu_{\rm c,a} \sim 10^{19}$ Hz and $\gamma^{2} \nu_{\rm m}
\equiv \nu_{\rm c,m} \sim 10^{19}$ Hz.

In figure 6, we show the allowed values of $n$ as a 
function of $r$ for the Synchrotron self-Compton (SSC) model. 
As in  figure 5, we found parameters of the allowed region: 
$n = 4.6 \times 10^{4}$ cm$^{-3}$, $r = 6.5 \times 10^{16}$ cm, 
$\gamma_{\rm m} \sim 260$, and $B = 16$ gauss, 
assuming $\xi = 1$ and $\epsilon_{\rm e} \sim 0.1$. 
Note that a $10^{16}$ cm radius is close to the typical value of the
``afterglow'' emission, rather than the prompt GRB emission 
(e.g., Piran 1999).
We therefore conclude that the SSC model for the hard X-ray 
spectrum is also unlikely in the standard  fireball scenario.

\begin{figure}
\begin{center}
\includegraphics[width=8cm,angle=270,keepaspectratio]{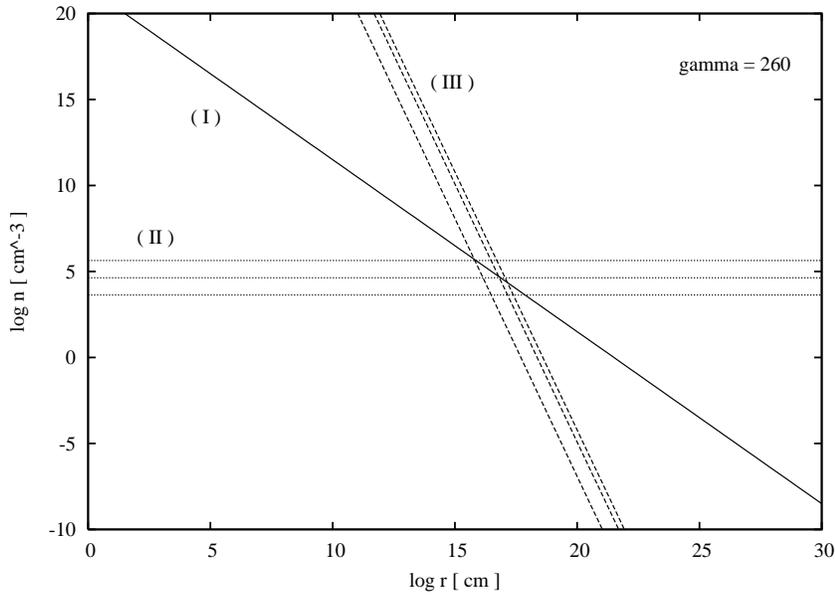}
\end{center}
\caption{Parameter space ($n, r$) allowed by the SSC model with
$\nu_{\rm c,a}$ of $10^{19}$ Hz, $\nu_{\rm c,m}$ of $10^{19}$ Hz,
and $E_{\rm total}$ of $5.8 \times 10^{51}$ ergs.
Lines (I), (II), and (III) represent the same relations as in figure 5.
The crossing point of the three expressions indicates the allowed parameters for the SSC model.}
\end{figure}

Thus, it is difficult to explain the P1 spectrum of GRB 020813 by 
either the SSA or SSC model. Therefore, we consider other 
radiation processes, such as ``jitter'' radiation and Compton drag effects. 
Medvedev (2000) pointed out that the radiation emitted by ultra-relativistic 
electrons in highly nonuniform, small-scale magnetic fields is different 
from synchrotron radiation if the 
electron's transverse deflections in these fields are much smaller 
than the beaming angle. He showed that the spectral power distribution 
of the radiation produced by  the power-law distributed electrons 
is well described by a sharply broken power-law, 
$\propto \nu^1$ for $\nu \le \nu_{\rm jb}$ and 
$\propto \nu^{-(p-1)/2}$ for $\nu \ge \nu_{\rm jb}$, 
where $p$ is the electron power-law index and $\nu_{\rm jb}$ is 
the jitter break frequency (Medvedev 2000).

Therefore, the low-energy depletion ($\alpha$ is larger than the death-line) 
with a spectral break may be naturally explained by jitter radiation.
In this model, the spectral break due to jitter will be observed at 

\begin{equation}
\nu_{\rm m} \sim 6.0 \times 10^{15} 
\biggl(\frac{\gamma_{\rm m}}{100} \biggr)^{3}
\biggl(\frac{\gamma_{\rm int}}{1} \biggr)
\biggl(\frac{\gamma_{\rm e}}{1} \biggr)^{-1/2}
\biggl(\frac{n}{10^{10}} \biggr)^{1/2}
\hspace*{0.2cm} {\rm Hz},
\end{equation}
where  $\gamma_{\rm int}$ is the relative Lorentz factor of two 
colliding shells, $\gamma_{\rm int} \sim 2-4$, and  
$\gamma_{\rm e}$ is the initial effective thermal Lorentz factor, 
$\gamma_{\rm e} \sim 2-3$ (Medvedev 2000). If  
$\gamma_{\rm m}$ is close to 1000, $\nu_{\rm m}$ can be observed in 
the X-ray band. In the realistic case, the emergent spectrum is 
determined by the statistical properties of the magnetic field in the 
emission region. If the magnetic field is highly inhomogeneous, 
jitter radiation may overcome the synchrotron radiation, whereas  
the reverse applies for the uniform magnetic field cases. Therefore, 
it is possible that the X-ray spectrum of GRB 020813
(P1 region) may be a mixture of jitter radiation with the usual 
synchrotron radiation component, but detailed modeling of the 
spectrum is beyond the scope of this paper.

As an alternative idea, Lazzati et al. (1999) and Ghisellini et al. (2000) 
proposed that the gamma-ray photons that characterize the prompt 
emission of GRBs are produced through the Compton drag process, caused 
by the interaction of a relativistic fireball with a very dense soft 
photon bath. This is an interesting assumption, because some of the 
GRBs are indeed associated with supernovae, where the expanding star 
can provide enough soft photons to make the radiative drag effective.

This model accounts for the basic properties of GRBs, 
i.e., the overall energetics and the peak frequency of the spectrum, 
with a radiative efficiency of more than 50$\%$. Also note that the 
GRB  should have a very hard spectrum $\propto \nu^2$, as 
expected from the Rayleigh-Jeans law below the peak frequency of 
the emission. 
In fact, our spectral shape is in good agreement with the spectra 
produced by Compton drag for $\Gamma_{0} = 300$ (Ghisellini et al. 2000).
The major weakness of this model is that the predicted spectrum is
basically thermal, and a typical power-law spectrum is not 
always produced. 
Thus, we find that both the jitter radiation and the Compton drag 
may reproduce the low-energy depletion observed in GRB 020813.

However, we note that a complete theoretical interpretation of the data 
is beyond the scope of this paper. 
There are other theoretical models that can explain the spectra with low-energy 
depletion.

\subsection{Spectrum in P2}
 
The average P2 spectrum is well described by the Band function with 
$\alpha \sim -1.00$ and $\beta \sim -1.76$. In contrast to P1,
$\alpha$ falls in the range allowed by the standard synchrotron shock model (SSM).
As shown by Sakamoto et al. (2005), integration of the
BATSE and the HETE-2 data revealed that the ``typical'' GRB
spectrum is well represented by a broken power-law (Band) function
with $\alpha \sim -1$ and $\beta \sim -2.5$.
Therefore, the observed break energy corresponds
to the peak energy ($\nu \sim \nu_{\rm m}$) in $\nu{\rm F}_\nu$
space, which is where most of the GRB power is emitted.
For the case of P2 in GRB 020813, however, the spectrum is 
still $rising$ up to $>$ 400 keV. 
\footnote{Since the detection limit of HETE-2 is 400 keV, 
it is difficult to discuss the slope of the high-energy power-law above 400 keV. 
It is necessary to consider the results from a higher energy detector such as Konus.} 
In the standard picture of SSM, 
such a break could be explained only when $\nu_{\rm m}$ is
well above the HETE bandpass, which again seems to be
quite different from the P1 spectrum.
  
We stress, however, that  it does not seem to be strange even if P1 and P2 show
quite different spectral features, since they  constitute  distinct
``peaks'' both in the light curve and in the hardness ratio (figure 4).
According to the internal shock scenario (e.g., Piran 1999), 
GRB light curves are thought to be produced by the collisions of shells 
that are moving with different Lorentz factors.  A fraction of the bulk 
kinetic energy is converted to the random energy of electrons, which emit the 
$observed$ radiation as prompt GRB emission. Since the radiation from 
each two-shell collision would be observed as a single pulse, P1 and P2 may be 
attributed to the collisions of different pairs of shells, 
which are possibly taking place in different physical conditions 
(e.g., magnetic field, density of material).
 
A detailed simulation of a collisionless shock plasma predicts that the magnetic field 
produced in GRB shocks randomly fluctuates on a very small scale of 
the relativistic skin depth, which is $\lambda\sim 10^2$ cm in internal shocks 
(Medvedev, Loeb 1999). Meanwhile, the emitting ultrarelativisitic electrons have 
much larger Larmor radii. In such situations, jitter radiation may overcome 
the usual synchrotron radiation, but it is also likely that the large scale fields 
are also present in the GRB ejecta. Therefore, the dominance of the ``synchrotron'' 
and the ``jitter'' radiation would be determined by a combination of large-scale fields 
($B_{\rm LS}$) and non-uniform, small-scale magnetic fields ($B_{\rm SS}$).  
We mention in this context that the different observed properties in P1 and P2 
may reflect the different balance of large/small-scale magnetic fields, 
where  $B_{\rm SS} \ge B_{\rm LS}$ for P1 and $B_{\rm SS} \ge B_{\rm LS}$ for P1 
and $B_{\rm SS} \le B_{\rm LS}$ for P2, but further study using more data 
is necessary to confirm this.

\section{Conclusion}

In this paper we reported HETE-2 WXM/FREGATE observations of 
bright and long GRB 020813. 
Our observational results demonstrate that the peak P1 (17$-$52 s) of the burst 
shows a depletion of low-energy photons below about 50 keV.
Therefore, we examined various scenarios to understand the origin of 
such a low-energy depletion.
We conclude that it is difficult to explain the depletion by either 
the SSA or SSC model. 
One possibility is that the low-energy depletion may be understood as 
a mixture of ``jitter'' radiation with the usual synchrotron radiation 
component. 
On the other hand, the spectrum of P2 (67$-$109 s) can 
be explained by the standard SSM. We believe that the magnetic field is uniform 
and that synchrotron radiation may overcome the jitter radiation in this region.   
\\
We are grateful to R. Yamazaki for giving useful advise concerning our discussions. 
The HETE-2 mission is supported in the US by NASA contract NASW-4690; 
in Japan in part by Grant-in-Aid (14079102 and 14GS0211) from 
the Ministry of Education, Culture, Sports, Science, and Technology; 
and in France by CNES contract 793-01-8479.  K.H. is grateful for support
under MIT-SC-R-293291. 

%%%
% See the manual for the detail.
%%%

\end{document}